\documentclass[conference]{IEEEtran}
\IEEEoverridecommandlockouts
\usepackage{cite}
\usepackage{amsmath,amssymb,amsfonts}
\usepackage{algorithmic}
\usepackage{graphicx}
\usepackage{textcomp}
\usepackage{xcolor}
\def\BibTeX{{\rm B\kern-.05em{\sc i\kern-.025em b}\kern-.08em
    T\kern-.1667em\lower.7ex\hbox{E}\kern-.125emX}}

\usepackage{microtype}
\usepackage{graphicx}
\usepackage{subfigure}
\usepackage{booktabs} 
\usepackage{mathtools}
\usepackage{amsthm}
\usepackage{algorithm}
\usepackage{algorithmic}
\usepackage[capitalize,noabbrev]{cleveref}

\theoremstyle{plain}

\theoremstyle{definition}

\theoremstyle{remark}

\usepackage{bm}
\usepackage{bbm}
\usepackage{comment}         
\usepackage{multicol}
\usepackage{multirow}
\usepackage{caption}
\usepackage{natbib}
\usepackage{braket}
\usepackage{enumitem}

\DeclareMathOperator*{\argmax}{arg\,max}
\DeclareMathOperator*{\argmin}{arg\,min}

\allowdisplaybreaks

\begin{document}

\title{Bayesian Portfolio Optimization\\
by Predictive Synthesis}

\author{\IEEEauthorblockN{Masahiro Kato}
\IEEEauthorblockA{\textit{Data Analytics Department} \\
\textit{Mizuho-DL Financial Technology Co., Ltd.}\\
Tokyo, Japan \\
masahiro-kato@fintec.co.jp}
\and
\IEEEauthorblockN{Kentaro Baba}
\IEEEauthorblockA{\textit{Data Analytics Department} \\
\textit{Mizuho-DL Financial Technology Co., Ltd.}\\
Tokyo, Japan \\
kentaro-baba@fintec.co.jp}
\and
\IEEEauthorblockN{Hibiki Kaibuchi}
\IEEEauthorblockA{\textit{Data Analytics Department} \\
\textit{Mizuho-DL Financial Technology Co., Ltd.}\\
Tokyo, Japan \\
hibiki-kaibuchi@fintec.co.jp
}
\and
\IEEEauthorblockN{Ryo Inokuchi}
\IEEEauthorblockA{\textit{Data Analytics Department} \\
\textit{Mizuho-DL Financial Technology Co., Ltd.}\\
Tokyo, Japan \\
ryo-inokuchi@fintec.co.jp}
}

\maketitle

\begin{abstract}
Portfolio optimization is a critical task in investment. Most existing portfolio optimization methods require information on the distribution of returns of the assets that make up the portfolio. However, such distribution information is usually unknown to investors. Various methods have been proposed to estimate distribution information, but their accuracy greatly depends on the uncertainty of the financial markets. Due to this uncertainty, a model that could well predict the distribution information at one point in time may perform less accurately compared to another model at a different time. To solve this problem, we investigate a method for portfolio optimization based on Bayesian predictive synthesis (BPS), one of the Bayesian ensemble methods for meta-learning. We assume that investors have access to multiple asset return prediction models. By using BPS with dynamic linear models to combine these predictions, we can obtain a Bayesian predictive posterior about the mean rewards of assets that accommodate the uncertainty of the financial markets. In this study, we examine how to construct mean-variance portfolios and quantile-based portfolios based on the predicted distribution information.
\end{abstract}

\begin{IEEEkeywords}
Portfolio optimization, Bayesian analysis, ensemble algorithm
\end{IEEEkeywords}

\section{Introduction}
Portfolio optimization is a critical challenge in investment, where the goal is to hold multiple financial assets in an appropriate allocation to achieve desirable asset management for investors. 

There are several approaches to understanding the risk of a portfolio. The mean-variance approach, one of the most classical criteria, was proposed by Markowitz and is also known as the Markowitz portfolio \citep{Markowitz1952,Markowitz1959,Markowitz2000}. In the mean-variance approach, the portfolio's variance is considered as the risk, and financial assets are allocated considering the trade-off between the portfolio's expected value and variance. The quantile-based approach is also influential, where Value at Risk (VaR) and Conditional VaR (CVaR) are used as risk metrics. \cite{Rockafellar2000OptimizationOC} proposed constructing a portfolio by minimizing CVaR using linear programming. Moreover, the risk parity approach is a favored approach among investors, where assets are allocated so that the ratios of the variances of the financial assets' returns become the portfolio's risk.

Optimization of portfolios based on these criteria often requires information about the distribution of asset returns. For example, constructing a mean-variance portfolio requires the input of estimated means and variances of asset returns. Similarly, the quantile-based approach needs the shape of the distribution, and the risk parity approach requires the covariance matrix. This information on distributions is usually unknown to investors and needs to be estimated from data. Since the construction of portfolios depends on the input distribution information, estimation errors in this information can affect the portfolio composition and sometimes significantly degrade its performance \cite{Chopra1993}.

Many difficulties in estimating information about the distribution of asset returns from data are due to market uncertainties. When the target time series is non-stationary or access is limited to data of a small sample size, the difficulty of estimating information about the distribution increases, making it challenging to construct portfolios.

This study adopts a Bayesian approach to tackle this issue. Firstly, we assume the existence of experts with their own predictions of asset returns. Then, we use Bayesian Predictive Synthesis (BPS), one of the Bayesian ensemble methods, to integrate these predictions \citep{mcalinn2019dynamic,mcalinn2020multivariate}. Bayesian Predictive Synthesis is a general framework that includes Bayesian model averaging as a special case. Following \cite{mcalinn2019dynamic} and \cite{mcalinn2020multivariate}, this paper uses dynamic linear models, which are considered suitable for time series prediction. As a result of BPS, we obtain a predictive distribution for each asset return. Under this predictive distribution, we can evaluate the portfolio under each weight $\bm{w}$. By optimizing the evaluation value for $\bm{w}$, we can select appropriate weights. Among various criteria for weight selection, we consider the mean-variance approach, the quantile-based approach, and the risk parity approach.

The contribution of this study lies in investigating the outcomes when using practically significant portfolio selection criteria under the posterior predictive distribution obtained by BPS. The BPS of mean-variance portfolios has been explored from the perspective of quadratic utility maximization by \cite{tallman2023bayesian}. This study further empirically considers the use of constrained optimization. Moreover, to our knowledge, the use of BPS in the quantile-based and risk parity approaches has not been thoroughly investigated. This study examines what outcomes can be obtained when using BPS for such portfolio construction methods.

\section{Problem Setting}
In this section, we formalize the problem of portfolio optimization. We consider optimizing a portfolio consisting of $K$ types of financial assets over $T$ periods, allowing for changes in the portfolio's composition.

\subsection{Asset Returns}
Let $X_{a, t} \in \mathbb{R}$ be the return of a financial asset $a \in [K] \coloneqq \{1, 2, \dots, K\}$ in period $t$. The return vector for $K$ types of financial assets is denoted as
\[\bm{X}_t = \big(X_{1, t}, X_{2, t}, \dots, X_{K, t}\big)^\top.\]
Here, let $\bm{x}_t$ be the realized value of $\bm{X}_t$. Also, let $\bm{X}_{s:t} = \{\bm{X}_s, \bm{X}_{s+1}, \dots, \bm{X}_t\}$ be the set of asset returns from period $s$ to $t$, with its realized values denoted as $\bm{x}_{s:t}$.

\subsection{Portfolio}
Define the set of portfolio weights as $\Delta^K \coloneqq \{z \in [0, 1]^K \mid \sum^K_{i=1} z_i = 1\}$. For simplicity, short selling is not allowed. Investors hold assets based on certain weights $\bm{w} \in \Delta^K$ and receive their returns. The return of a portfolio under the weights $\bm{w} \in \Delta^K$ can be written as
\begin{align*}
    R_t(\bm{w}) = \bm{w}^\top\bm{X}_t.
\end{align*}
We construct a portfolio by choosing a desirable $\bm{w}$ under a suitable criterion. For simplicity, we assume that changing the portfolio's composition at each time does not incur any costs.

\subsection{Portfolio Selection Criteria}
In this study, we consider constructing portfolios under the Bayesian posterior predictive distribution obtained through BPS. Specifically, we focus on three approaches: mean-variance portfolios, quantile-based portfolios, and risk-parity portfolios, using the posterior predictive distribution.

\section{BPS}
This section discusses obtaining the posterior predictive distribution of asset returns based on BPS. The method follows \cite{mcalinn2020multivariate}.

\subsection{Experts}
To construct a portfolio, we need to input information about the distribution of asset returns $\bm{X}_t$. In this study, we assume the existence of $J$ experts who provide predictive distributions for the mean of $\bm{X}_t$. Investors can construct portfolios based on these predictive distributions.

Let the state about the assets' mean rewards prediction of an expert $j \in [J]$ at time $t$ be denoted by the $K$-dimensional vector $\bm{z}_{tj} \coloneqq \big(z_{t1j}, z_{t2j}, \dots, z_{tKj}\big)^\top$. In this paper, each state $z_{taj}$ represents the prediction of expert $j \in [J]$ for the price of asset $a \in [K]$ in period $t$. Also, let
\[\bm{z}_t \coloneqq \big(\bm{z}_{t1}, \dots, \bm{z}_{tJ}\big).\]

The predictive distribution of the state $\bm{z}_{tj}$ of expert $j$ is denoted by $h_{tj}(\bm{z})$. Let the set of predictive distributions for each asset and each expert in period $t$ be $H_t \coloneqq (h_{tj})_{j \in [J]}$. Also, let the set of predictive distributions up to period $t$ be $H_{1:t} = (H_s)^t_{s=1}$.

BPS is a method for integrating these experts' predictive distributions to construct a new predictive distribution.

\subsection{Modeling Portfolio Returns}
\label{sec:bayesian_model}
Assume that the asset return vector follows a multivariate normal distribution
\begin{align*}
    \bm{X}_t \mid \bm{\mu}, \Sigma \sim \mathcal{N}\big(\bm{\mu}, \Sigma\big),
\end{align*}
where $\mathcal{N}\big(\bm{\mu}, \Sigma\big)$ is a multivariate normal distribution with a mean vector $\bm{\mu}$ and a covariance matrix $\Sigma$. 
Under this assumption, the return of the portfolio follows a normal distribution
\begin{align*}
    R_t(\bm{w}) \mid \bm{\mu}, \Sigma \sim \mathcal{N}\big(\bm{w}^\top\bm{\mu}, \bm{w}^\top\Sigma\bm{w}\big).
\end{align*}

Let $\bm{\mu}_t$ and $\Sigma_t$ denote $\bm{\mu}$ and $\Sigma$ given by posterior samples, respectively. Then, the posterior density of the portfolio return $p\big(r_t(\bm{w}) \mid \bm{x}_{1:(t-1)}, \mathcal{H}_{1:t}\big)$ under the observation of asset returns and the predictive model $\mathcal{H}_{1:t}$ of experts is given by
\begin{align}
\label{eq:post}
&p\big(r_t(\bm{w}) \mid \bm{x}_{1:(t-1)}, \mathcal{H}_{1:t}\big)\\
&\coloneqq \int  p\big(r_t(\bm{w}) \mid \bm{\mu}_t, \Sigma_t, \bm{x}_{1:(t-1)}, \mathcal{H}_{1:t}\big)\nonumber\\
&\ \ \ \ \ \ \ \ \ \ \ \ \ \ \ \ \ \ \times p\big(\bm{\mu}_t, \Sigma_t \mid \bm{x}_{1:(t-1)}, \mathcal{H}_{1:t}\big)\mathrm{d}\Phi_t.\nonumber
\end{align}

\subsection{Synthesis Function}
From \eqref{eq:post}, to calculate the posterior density $p\big(r_t(\bm{w}) \mid \bm{x}_{1:(t-1)}, \mathcal{H}_{1:t}\big)$, it is sufficient to specify $p\big(r_t(\bm{w}) \mid \bm{\mu}_t, \Sigma_t, \bm{x}_{1:(t-1)}, \mathcal{H}_{1:t}\big)$ and $p\big(\bm{\mu}_t, \Sigma_t \mid \bm{x}_{1:(t-1)}, \mathcal{H}_{1:t}\big)$.

First, we model $p\big(r_t(\bm{w}) \mid \bm{\mu}_t, \Sigma_t, \bm{x}_{1:(t-1)}, \mathcal{H}_{1:t}\big)$. We consider a model where the effects of observed data $\bm{x}_{1:(t-1)}$ are only reflected in the parameters $\bm{\mu}_t$ and $\Sigma_t$. That is,
\[p\big(r_t(\bm{w}) \mid \bm{\mu}_t, \Sigma_t, \bm{x}_{1:(t-1)}, \mathcal{H}_{1:t}\big) = p\big(r_t(\bm{w}) \mid \bm{\mu}_t, \Sigma_t, \mathcal{H}_{1:t}\big).\]

As addressed in this study, there is uncertainty in estimating the parameters $\bm{\mu}_t$ and $\Sigma_t$. We consider obtaining these parameters by integrating the predictive distributions of experts represented by $H_{1:t}$, as
\begin{align*}
&p\big(r_t(\bm{w}) \mid \bm{\mu}_t, \Sigma_t, \mathcal{H}_{1:t}\big)\\
&\coloneqq \int \alpha_t\big(r_t(\bm{w}) \mid \bm{z}_t, \bm{\mu}_t, \Sigma_t\big)\prod_{j \in [J]}h_{tj}(\bm{z}_{t, j})\mathrm{d}\bm{z}_{t, j},
\end{align*}
where $\alpha_t: \mathbb{R} \times \mathbb{R} \times \mathbb{R} \times \mathbb{R}^{K \times K} \to \mathbb{R}$ is called the \emph{synthesis function}. By changing the definition of this synthesis function, various models can be treated as a form of BPS. For example, Bayesian model averaging is included as a special case \cite{hoeting1999bayesian,geweke2011optimal,aastveit2018combined}.

\paragraph{Dynamic Linear Models}
Various definitions can be given to the synthesis function $\alpha_t$, but in this study, we focus on \emph{dynamic linear models} following \cite{mcalinn2019dynamic} and \cite{mcalinn2020multivariate}, as
\begin{align}
    &\bm{X}_t = \bm{\mu}_t + \bm{\nu}_t, \qquad \bm{\nu}_t \sim \mathcal{N}(0, \bm{V}_t),\nonumber\\
    &\bm{\mu}_t = F(\bm{z}_t)\bm{\beta}_t,\nonumber\\
    \label{eq:time_varying_beta}
    &\bm{\beta}_t = \bm{\beta}_{t-1} + \bm{\omega}_t, \qquad \bm{\omega}_t \sim \mathcal{N}(0, \bm{W}_t).
\end{align}
Here,
\begin{align*}
    F(\bm{z}_t) \coloneqq \begin{pmatrix}
        1 & \bm{f}^\top_{t1} & 0 & \bm{0}^\top & \cdots & \cdots 0 & \bm{0}^\top\\
        0 & \bm{0}^\top & 1 & \bm{f}^\top_{t2} & & & \vdots\\
        \vdots & & & & \ddots & & \vdots \\
        0 & \bm{0}^\top & \cdots & \cdots & \cdots & 1 & \bm{f}^\top_{tK}
    \end{pmatrix},
\end{align*}
where $\bm{f}^\top_{tk} \coloneqq (z_{tk1}, z_{tk2}, \dots, z_{tkJ})$ is a $J \times 1$ vector representing the predictions of $J$ experts for the return of asset $k$. Also, $\bm{\beta}_t$ is a $(J+1) \times K$ vector. Recall that $z_{tj} = \big(\bm{z}_{t1}, \dots, \bm{z}_{tJ}\big)$ is generated from $h_{tj}(z_{tj})$.

This model is a type of state space model and is considered suitable for modeling time series data, as addressed in this study. Let the set of parameters of the dynamic linear model be $\Phi_t \coloneqq \big(\bm{\beta}_t, \bm{V}_t, \bm{W}_t\big)$. Then, the synthesis function can be rewritten as
\begin{align*}
    \alpha_t\big(r_t(\bm{w}) \mid \bm{z}_t, \bm{\mu}_t, \Sigma_t\big) = \alpha_t\big(r_t(\bm{w}) \mid \bm{z}_t, \Phi_t\big).
\end{align*}

Under this dynamic linear model, the time-varying coefficient $\bm{\beta}_t$ follows a random walk defined by \eqref{eq:time_varying_beta}. Here,
$\bm{W}_t$ is defined via a standard single discount factor specification~(Section~6.3 in \cite{WestHarrison1997book2}; Section~4.3 in \cite{Prado2010}), using a state evolution discount factor $e\in (0,1]$. Moreover, the residual variance $\varepsilon_t$ follows a standard beta-gamma random walk volatility model (Section~10.8 in \cite{WestHarrison1997book2}; Section~4.3 in \cite{Prado2010}), with $\varepsilon_t = \varepsilon_{t-1}\delta/\gamma_t$ for some discount factor $\delta\in (0,1]$ and where $\gamma_t$ are beta distributed innovations, independent over time and independent of $\nu_s$ and $\eta_{1,r},\dots, \eta_{J,r}$ for all $t,s,r$. Given choices of discount factors underlying these two components, and a (conjugate normal/inverse-gamma) prior for $(w_{0,0},w_{1,0},\ldots,w_{J,0}, \nu_0)$ at $t=0,$ the model is specified.

\subsection{Posterior Predictive Distribution}
As a result of this Bayesian modeling, we can obtain the posterior predictive distribution as
\begin{align*}
&p\big(r_t(\bm{w}) \mid \bm{x}_{1:(t-1)}, \mathcal{H}_{1:t}\big) \coloneqq\\
&\int  p\big(r_t(\bm{w}) \mid \bm{x}_{1:(t-1)}, \Phi_t, \mathcal{H}_{1:t}\big) p\big(\Phi_t \mid \bm{x}_{1:(t-1)}, \mathcal{H}_{1:t}\big)d\Phi_t,
\end{align*}
where
\begin{align*}
&p\big(r_t(\bm{w}) \mid \bm{x}_{1:(t-1)}, \Phi_{t}, \mathcal{H}_{1:t}\big) \coloneqq p\big(r_t(\bm{w}) \mid \Phi_{t}, \mathcal{H}_{1:t}\big)\\
&= \int \alpha_t\big(r_t(\bm{w}) \mid \bm{z}_t, \Phi_t\big)\prod_{j \in [J]}h_{tj}(\bm{z}_{t, j})\mathrm{d}\bm{z}_{t, j}.
\end{align*}

We can obtain the information required for portfolio optimization from this posterior predictive distribution. For example, the expectation of some function $g: \mathbb{R} \to \mathbb{R}$ can be calculated as
\begin{align*}
&\mathbb{E}\left[g(R_t(\bm{w})) \mid \bm{x}_{1:(t-1)}, \mathcal{H}_{1:t}\right]\\
&= \int \int  \int g(r_t(\bm{w}))\alpha_t\big(r_t \mid \bm{z}_t, \Phi_t\big)\nonumber\\
&\ \ \ \prod_{j \in [J]}h_{tj}(z_{t, j}) p\big(\Phi_t \mid \bm{x}_{1:(t-1)}, \mathcal{H}_{1:t}\big)\mathrm{d}r_t(\bm{w})\mathrm{d}z_{t, j}\mathrm{d}\Phi_t\nonumber.
\end{align*}

In BPS, since the posterior distribution cannot be obtained analytically, it is computed by simulation using Markov Chain Monte Carlo (MCMC). The details of MCMC are described in \cite{mcalinn2020multivariate}.

\section{Bayesian Portfolio}
Here, we introduce portfolio optimization based on the predictive distribution obtained through BPS.

\subsection{Mean-Variance Portfolio}
First, we discuss the mean-variance approach based on the Bayesian posterior predictive distribution. \cite{Bauder2021} proposes a method for constructing a portfolio independent of unknown parameters by expressing the parameters of the asset return distribution as a function of observed data under appropriate modeling. Additionally, \cite{tallman2023bayesian} proposes a mean-variance approach based on multivariate BPS. In this section, based on these prior studies, we examine the mean-variance approach utilizing BPS.

\paragraph{Constrained Optimization.} As a method to construct a mean-variance portfolio, we consider a constrained optimization problem characterized by the posterior mean $\mathbb{E}\left[R_t(\bm{w}) \mid \bm{x}_{1:(t-1)}, \mathcal{H}_{1:t}\right]$ and the posterior variance $\mathrm{Var}\left[R_t(\bm{w}) \mid \bm{x}_{1:(t-1)}, \mathcal{H}_{1:t}\right]$ at each period $t$, conditioned on $\bm{x}_{1:(t-1)}$ and $H_{1:t}$. Namely, the weights $\bm{w}^{\mathrm{MV}}$ of the mean-variance portfolio are defined as the solution to the optimization problem
\begin{align*}
    \bm{w}^{\mathrm{MV}} \in \argmin_{\bm{w} \in \Delta^K}&\ \ \ \mathrm{Var}\left[R_t(\bm{w}) \mid \bm{x}_{1:(t-1)}, \mathcal{H}_{1:t}\right]\\
    \mathrm{s.t.}&\ \ \ \mathbb{E}\left[R_t(\bm{w}) \mid \bm{x}_{1:(t-1)}, \mathcal{H}_{1:t}\right] \geq \eta,
\end{align*}
where $\eta > 0$ is the mean constraint.

\paragraph{Expected Quadratic Utility Maximization.}
The quadratic utility function of an investor operating a portfolio with weights $\bm{w} \in \Delta^K$ is defined as
\begin{align*}
    &U(\bm{w}) \coloneqq\\
    &\mathbb{E}\left[R_t(\bm{w}) \mid \bm{x}_{1:(t-1)}, \mathcal{H}_{1:t}\right] - \frac{\gamma}{2} \mathbb{E}\left[R_t^2(\bm{w}) \mid \bm{x}_{1:(t-1)}, \mathcal{H}_{1:t}\right],
\end{align*}
and the weights $\bm{w}$ of the mean-variance portfolio maximize the expected value of this quadratic utility. In Bayesian mean-variance portfolios, using the posterior mean, the weights $\widetilde{\bm{w}}^{\mathrm{MV}}$ of the mean-variance portfolio can be given as
\begin{align*}
    \widetilde{\bm{w}}^{\mathrm{MV}} \in \argmax_{\bm{w} \in \Delta^K}\mathbb{E}\left[U(\bm{w}) \mid \bm{x}_{1:(t-1)}, \mathcal{H}_{1:t}\right].
\end{align*}

The solution to the constrained optimization problem is known to correspond to the solution of this expected quadratic utility maximization problem under suitable conditions. \cite{tallman2023bayesian} in particular discusses the BPS-based mean-variance portfolio from the perspective of expected quadratic utility maximization.

\subsection{Quantile-Based Portfolio}
Next, we consider a quantile-based portfolio using the Bayesian posterior predictive distribution obtained through BPS. In this study, we adopt the Bayesian quantile-based risk metric defined by \cite{bodnar2020bayesian}.

\paragraph{VaR and CVaR.} Define a loss function $\ell: \mathbb{R} \to \mathbb{R}$ for the portfolio return $R(\bm{w})$, and denote $L(\bm{w}) \coloneqq \ell(R(\bm{w}))$. In this paper, we set $L(\bm{w}) = -R(\bm{w})$. Here, let $F_{L(\bm{w}), t-1}$ be the cumulative density function of $L(R(\bm{w}))$. Then, the Bayesian VaR that evaluates the loss incurred within a certain probability using the posterior distribution is defined as
\begin{align*}
    &\mathrm{VaR}_{\beta, t-1}\big(L(\bm{w})\big) \coloneqq \inf_{l \in \mathbb{R}}\Big\{F_{L(\bm{w}), t-1}(l) \geq \beta \Big\},
\end{align*}
where $\beta \in (0, 1)$ represents the quantile. Similarly, the CVaR, which represents the average loss when the portfolio return loss exceeds a certain probability level $\beta$, is defined as
\begin{align*}
    &\mathrm{CVaR}_{\beta, t-1}\big(L(\bm{w})\big)\\
    &\coloneqq \mathbb{E}\Big[L(\bm{w}) \mid L(\bm{w}) \geq \mathrm{VaR}_{\beta, t-1}(L(\bm{w})), \bm{x}_{1:(t-1)}, \mathcal{H}_{1:t}\Big].
\end{align*}
This definition of CVaR is based on \cite{Bodnar2021} and includes the result of Proposition 6 in \cite{Rockafeller2002} as a special case. \cite{bodnar2020bayesian} discusses methods for constructing portfolios using the Bayesian VaR and CVaR defined in this way.

\paragraph{Quantile-Based Portfolio on Returns.}
Extending the concepts of VaR and CVaR, we can also consider a portfolio based on the quantiles of returns. Similarly to VaR, define the Value-of-Return (VoR) as
\begin{align*}
    \mathrm{VoR}_{\alpha, t-1}\Big(R(\bm{w})\Big) \coloneqq \inf_{r \in \mathbb{R}}\Big\{F_{R(\bm{w}), t-1}(r) \geq \alpha \Big\},
\end{align*}
where $F_{R(\bm{w}), t-1}$ is the cumulative density function of $R(R(\bm{w}))$. Then, we define the Conditional VoR (CVoR) as
\begin{align*}
    \mathrm{CVoR}_{\alpha, t-1}\Big(R(\bm{w})\Big) \coloneqq \mathbb{E}\Big[R(\bm{w}) \mid R(\bm{w}) \geq \mathrm{VoR}_{\alpha}(R(\bm{w}))\Big].
\end{align*}

\paragraph{Portfolio Optimization.}
Following \cite{Bodnar2021}, we use VaR or VoR to obtain the portfolio weights $\bm{w}^{\mathrm{Q}}$ by solving
\begin{align}
\label{eq:quantial}
    &\bm{w}^{\mathrm{Q}} \in \max_{\bm{w} \in \Delta^K}\mathrm{CVoR}_{\alpha, t-1}(R(\bm{w}))\\
    &\mathrm{VaR}_{\beta, t-1}\big(L(\bm{w})\big) \leq v_0,\nonumber
\end{align}
where $v_0 \in \mathbb{R}$ is the maximum loss an investor is willing to bear under risk.

In addition, portfolio optimization in \cite{bodnar2020bayesian} considers an objective function defined as
\begin{align*}
    Q(\bm{w}) \coloneqq -R\big(\bm{w}\big) + q_{\alpha}\sqrt{\bm{w}^\top \Sigma_t \bm{w}},
\end{align*}
where $q_{\alpha}$ is an indicator depending on VaR or CVaR. For example, using the $\alpha$ quantile of the standard deviation $z_\alpha$ to relate to VaR, set $q_{\alpha} = z_{\alpha}$, and to relate to CVaR, set $q_{\alpha} = \frac{\exp\big(-z^2_{\alpha}/2\big)}{(1-\alpha)\sqrt{2\pi}}$. \cite{bodnar2020bayesian} learns weights by optimizing this objective function.

\subsection{Risk Parity Portfolio}
The weights of a risk parity portfolio are given so that the risk contributions of each asset comprising the portfolio are equal \cite{qian2006financial}. In this paper, we define a risk parity portfolio using the posterior predictive distribution. First, calculate the marginal risk contribution (MRC) of asset $a \in [K]$ to the portfolio as
\begin{align*}
    \mathrm{MRC}_a = \frac{1}{2}\frac{\partial \mathrm{Var}\big(R(\bm{w}) \mid \bm{x}_{1:(t-1)}, \mathcal{H}_{1:t}\big)}{\partial w_a} = \sum_{b \in [K]}\Sigma_{a, b, t-1}w_b,
\end{align*}
where $\Sigma_{a, b, t-1}$ is the variance-covariance matrix of the asset return's posterior distribution. Then, the risk contribution (RC) is
\begin{align*}
    \mathrm{RC}_a(\bm{w}) = w_a \mathrm{MRC}_a / \mathrm{Var}\big(R(\bm{w}) \mid \bm{x}_{1:(t-1)}, \mathcal{H}_{1:t}\big).
\end{align*}
A portfolio with equal risk contributions for each asset is called a risk parity portfolio. The weights $\bm{w}^{\mathrm{RP}}$ of a risk parity portfolio are obtained by solving
\begin{align*}
    \bm{w}^{\mathrm{RP}} \in \argmin_{\bm{w} \in \Delta^K} \sum_{a \in [K]}\sum_{b \in [K]}\left(\mathrm{RC}_a(\bm{w}) - \mathrm{RC}_b(\bm{w})\right)^2.
\end{align*}

\section{Experiments}
In this study, we construct two empirical studies in the US and Japanese markets. In each market, we use $10$ types of stocks listed in Tables~\ref{tab:us} and \ref{tab:jap}.

We use the stock prices of each company from January 1, 2008, to December 31, 2019. Returns are calculated monthly. Data from 2008 to 2010 is used only for learning the parameters, and the portfolio's performance is tested using data from 2011 to 2019. Parameter estimation continues sequentially after 2011. The reason for not using all data before 2011 is to allow the posterior distribution of BPS to converge in advance.

\begin{table}[t]
    \caption{US stock data}
    \label{tab:us}
    \centering
    \scalebox{0.85}{
    \begin{tabular}{|c|c|}
    \hline
    Company & Industry \\
    \hline
Apple Inc. & Technology \\ 
Microsoft Corp. & Technology \\ 
Amazon.com Inc. & Consumer Discretionary \\ 
Alphabet Inc. & Communication Services \\ 
Berkshire Hathaway Inc. & Financials (Diversified Holdings) \\ 
Johnson \& Johnson & Health Care \\ 
Walmart Inc. & Consumer Staples (Retail) \\ 
ExxonMobil Corp. & Energy (Oil and Gas) \\ 
Procter \& Gamble Co. & Consumer Staples (Consumer Goods) \\
Intel Corp. & Technology (Semiconductors) \\ 
\hline
    \end{tabular}
}
\end{table}

\begin{table}[t]
    \caption{Japanese stock data}
    \label{tab:jap}
    \centering
    \scalebox{0.85}{
    \begin{tabular}{|c|c|}
    \hline
    Company & Industry \\
    \hline
Toyota Motor & Automotive \\
SoftBank Group & Telecommunication \& IT \\
Keyence & Electronic Equipment \\ 
Nidec Corporation & Electrical Equipment \\
Nintendo & Entertainment \\ 
Tokyo Electron & Semiconductor Manufacturing Equipment \\ 
Fast Retailing & Retail (Apparel) \\ 
Tokio Marine Holdings & Insurance \\
Astellas Pharma & Pharmaceuticals \\ 
Seven \& i Holdings & Retail (General) \\
\hline
    \end{tabular}
}
\end{table}

\begin{figure*}[h]\centering
\includegraphics[width=180mm]{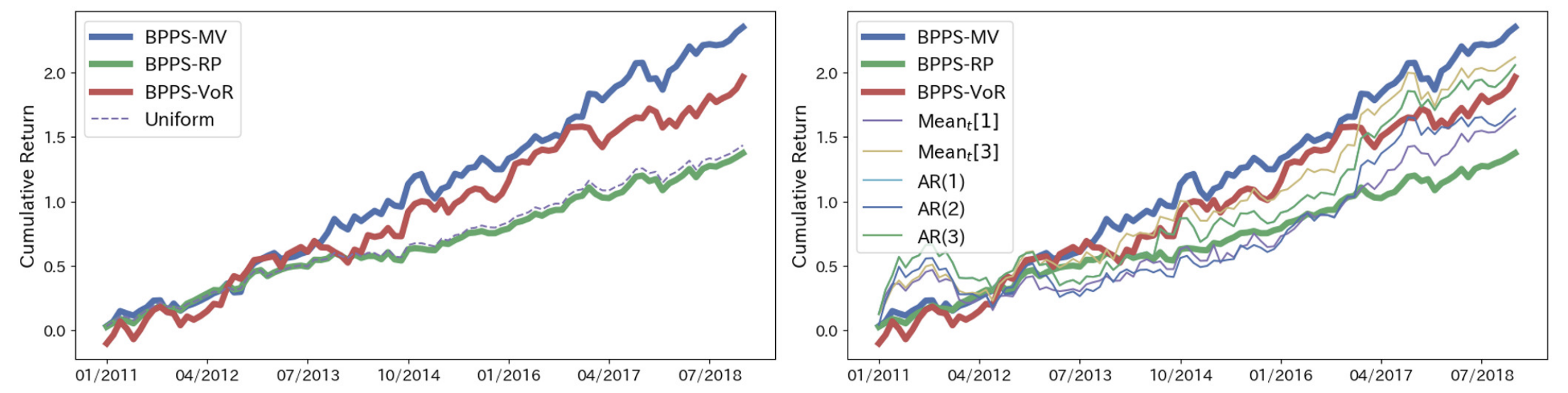}
\caption{Experimental results with US stocks. The $y$-axis in the figures represents the cumulative returns, while the $x$-axis represents the months and years. The left figure compares the proposed method with the equally weighted portfolio (denoted as Uniform), and the right figure compares the proposed method with the results obtained using sample means and AR models to predict returns.}
\label{fig:fig1}
\end{figure*}

\begin{figure*}[h]\centering
\includegraphics[width=180mm]{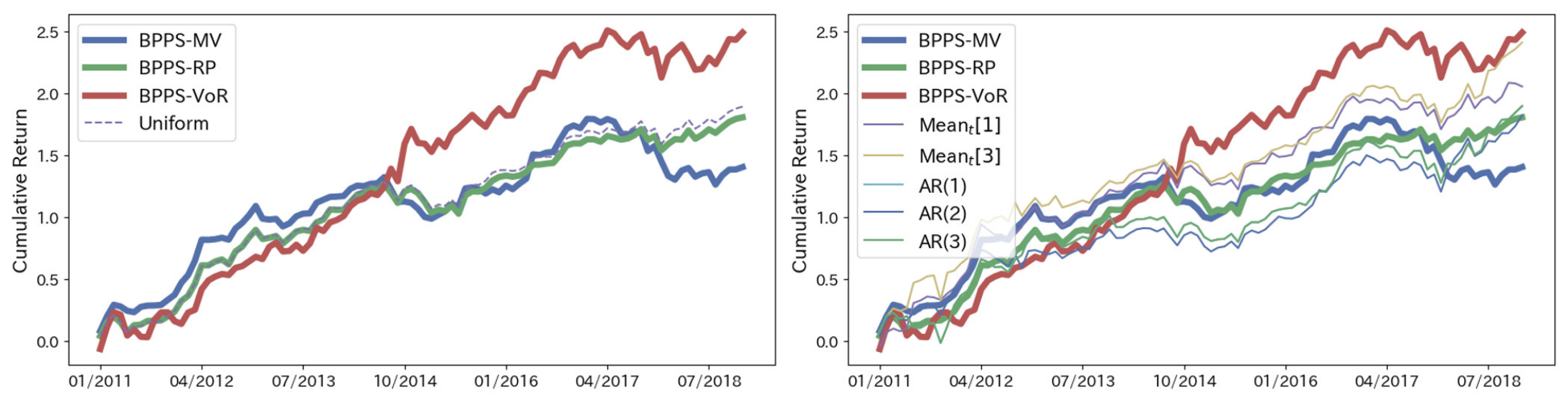}
\caption{Experimental results with Japanese stocks. The $y$-axis in the figures represents the cumulative returns, while the $x$-axis represents the months and years. The left figure compares the proposed method with the equally weighted portfolio (denoted as Uniform), and the right figure compares the proposed method with the results obtained using sample means and AR models to predict returns.}
\label{fig:fig2}
\end{figure*}

\subsection{Experts}
In BPS, multiple predictive models for asset returns $\bm{X}_t$ are treated as experts, and their predictions are integrated. In this paper, $\bm{X}_t$ is predicted using the following methods:
\begin{itemize}
    \item The sample mean of the past 1 year ($Mean_t[1]$).
    \item The sample mean of the past 3 years ($Mean_t[3]$).
    \item An AR$(1)$ regression model using samples from the past 3 years ($AR_t(1)$).
    \item An AR$(2)$ regression model using samples from the past 3 years ($AR_t(2)$).
    \item An AR$(3)$ regression model using samples from the past 3 years ($AR_t(3)$).
\end{itemize}

\subsection{Portfolio Construction Methods}
In this experiment, in addition to the mean-variance portfolio, the quantile-based (VoR) portfolio, and the risk-parity portfolio mentioned above, we use an equally weighted portfolio (setting $w_1 = \cdots = w_K = \frac{1}{K}$, denoted as Uniform) to test the performance of the portfolios. Furthermore, we also investigate the results when replacing the parameters with those estimated not by the Bayesian posterior distribution but by the sample means and AR models mentioned above. The Bayesian portfolio construction method based on BPS is denoted as BPPS (Bayesian Portfolio optimization by Predictive Synthesis). The BPPS based on the mean-variance portfolio is denoted as BPPS-MV, the BPPS based on the quantile portfolio is denoted as BPPS-VoR, and the BPPS based on the risk-parity portfolio is denoted as BPPS-RP.

\textbf{BPPS-MV.} In BPPS-MV, we construct the mean-variance efficient portfolios and then choose a portfolio with the highest Sharpe ratio.

\textbf{BPPS-VoR.} In BPPS-VoR, we construct portfolios by solving the constrained problem in \eqref{eq:quantial}. We set $\alpha = 0.05$, $\beta = 0.95$, and $v_0 = -0.1$. The loss function is the negative of the return. We solve the constrained problem by adding the penalty for violating the constraint to the objective as $\max_{\bm{w} \in \Delta^K}\big\{\mathrm{CVoR}_{\alpha, t-1}(R(\bm{w})) - \lambda \max\big\{0, \mathrm{VaR}_{\beta, t-1}\big(L(\bm{w})\big) - v_0\big\}\big\}$, where we set $\lambda = 10$.

\subsection{Experimental Results}
The experiments report the cumulative returns when operating the portfolio from January 1, 2008, to December 31, 2019. It is assumed that the portfolio's composition can be changed monthly and that there are no costs associated with these changes.

We show the results with US stocks in Figure~\ref{fig:fig1} and those with Japanese stocks in Figure~\ref{fig:fig2}.

In each of Figure~\ref{fig:fig1} and Figure~\ref{fig:fig2}, the left figure compares the proposed method with the equally weighted portfolio (Uniform), while the right figure compares the proposed method with the results when using sample means and AR models to predict returns. In the right figure, a mean-variance portfolio is used when using sample means and AR models. In that case, the variance is calculated using the variance of returns from the past 3 years. For all mean-variance portfolios, the portfolio on the efficient frontier with the highest Sharpe ratio is selected.

The experimental results show that BPPS performs well overall during the evaluation period without significant drops in performance. Although BPPS-MV experiences a significant performance drop towards the end in the Japanese market, it otherwise demonstrated higher performance than existing single prediction models or performed comparably to the best model among them. It is notable that our algorithm, despite using some models with empirically poor performance, minimally feels the impact of these inferior models, indicating the robustness of the BPPS approach.

In the US market, both BPPS-MV and BPPS-VoR show good performance. Remarkably, BPPS-VoR demonstrates the best performance and maintained high stability.

Until around June 2017 in the Japanese market, the fact that BPPS does not significantly drop in returns compared to other methods suggests that the state transition of BPS functioned well. Interestingly, the performance of BPPS-MV and BPPS-VoR reverses between 2013 and 2014. Although BPPS continues to show good performance, the reversal indicates that there are state transitions that BPS cannot fully capture. Moreover, BPPS-MV shows good performance until around June 2017 but then experiences a significant drop in performance. We expect that solving these issues could further improve performance.

At least according to our results, BPPS-VoR is consistently showing high performance. We believe that using Bayesian algorithms to assess quantiles in the posterior distribution is well-suited for portfolio optimization. Thus, our proposed BPS-based algorithm not only provides practical performance but also offers academic insights.

\section{Conclusion}
This study introduced a method for optimizing portfolios based on the posterior predictive distribution obtained through BPS to address the uncertainty of the asset return distribution in portfolio optimization. By integrating the multiple experts' predictions of asset returns using dynamic linear models, we constructed predictive distributions that capture the uncertainty of time series data. Then, we developed mean-variance portfolios, quantile-based portfolios, and risk-parity portfolios utilizing the posterior predictive distribution. Through experiments using stock price data, we confirmed the effectiveness of the methods tested in this paper.

\bibliographystyle{plain}
\bibliography{BayesPortfolio.bbl}

\end{document}